 \documentclass[10pt,showpacs,preprintnumbers,aps,onecolumn]{revtex4}
\usepackage{amsmath}
\usepackage{graphicx}
\usepackage{dcolumn}
\usepackage{bm}
  \normalsize

\begin{document}
\title{ Dynamic Control of Collapse in a Vortex Airy Beam}

\author{Rui-Pin Chen$^{1,2}$, Khian-Hooi Chew$^{3}$, Sailing He$^{1,4}$}
\affiliation{$^{1}$Centre for Optical and Electromagnetic Research,
State Key Laboratory of Modern Optical Instrumentations,
JORCEP [KTH-ZJU-LU Joint Research Center of Photonics],
Zhejiang University (ZJU), Hangzhou 310058, China\\
$^{2}$School of Sciences, Zhejiang A $\&$ F University,
Lin'an 311300, Zhejiang Province, China\\
$^{3}$Department of Physics, Faculty of Science,
University of Malaya, Kuala Lumpur 50603, Malaysia\\
$^{4}$Department of Electromagnetic Engineering, School of Electrical Engineering,
Royal Institute of Technology (KTH), 100 44 Stockholm, Sweden}

\date{\today}

\begin{abstract}
We study the self-focusing dynamics and collapse of vortex Airy optical beams in a Kerr
medium. The collapse is suppressed compared to a non-
vortex Airy beam in a Kerr medium as a result of the existence of vortex fields.
The locations of collapse depend sensitively on the initial power, vortex order, and
modulation parameters. Unlike the collapses reported before for any beam, the
collapse may occur in a position where the initial field is nearly zero while no
collapse appears in the region where the initial field is mainly distributed.
This study sheds light on how to control and manipulate the
location of collapse based on the initial power,
vortex order and modulation parameter.
\end{abstract}


\maketitle

Nonlinear wave collapse has been investigated in many areas of physics, including optics,
fluidics, plasma physics, and Bose-Einstein condensates [1].
In nonlinear optics, collapses of
beams with various spatial distributions have been studied [1-3].
One of the challenges facing this field of interest
is the possibility of controlling and manipulating the collapse dynamics.
Recently, the Airy beam has attracted considerable attention after its experimental
generation by Siviloglou et al. [4] due to its intriguing properties and potential
applications such as weak-diffraction, transverse acceleration [4-6], self-healing [7],
and sorting microscopic particles [8]. The evolution characteristics of an Airy
beam in nonlinear medium have been studied [9-17], such as plasma channel generation [13],
laser filamentation [14], supercontinuum and solitary wave generation [15-17].
Vortices have been the subject of many studies and appear in many branches of physics [18].
A vortex Airy
beam is formed by superposition of an Airy beam and a vortex
optical field. Recently, interesting propagation dynamics and non-classicality
of a vortex Airy beam have been reported [19, 20].
The nonlinear dynamics of a vortex Airy
beam in a Kerr medium is also expected to give some interesting and nontrivial properties.

In this letter, we investigate the spatial collapse dynamics of a vortex Airy beam.
The coupling between the vortex and Airy beam strongly affects the nonlinear dynamic
properties of the vortex and Airy beam in the Kerr medium. The diffraction against
self-focusing of a vortex Airy beam is effectively enhanced compared to a non-vortex
Airy beam. In addition, the vortex of beam tends to suppress the collapse in a Kerr medium.
For a non-vortex Airy beam in a Kerr medium,
partial collapse can occur in the beam's center [12].
However, the collapse of the vortex Airy beam never occurs at the beam's center.
This is because the center of the vortex is located at the center of the beam.
Our findings reveal that the position
of the partial collapse and the propagation distance for the
appearance of partial collapse are dependent on the initial powers, vortex orders,
and modulation parameters. The partial collapse may occur in the main lobe, side lobes,
or outermost lobes, depending on the strength of initial powers,
the order of vortex and the modulation parameters.
The collapse can occur in the position where the initial field is almost zero
while no collapse appears in the position where the field originally exists with a appropriate power.
We further show that the initial power, vortex order and modulation parameters
can be exploited to control and manipulate the position of collapse dynamics
in a vortex Airy beam. Even if the initial power is ultrahigh,
the partial collapse occurs separately at the side lobes and
the beam still propagates along a similar accelerating curve
trajectory as that of an Airy beam in free space.
Since the field distribution of a
vortex Airy beam is modulated by some exponential factors,
we also study the effect of these exponential factors on the nonlinear evolution of the beam in a Kerr medium.
Finally, the evolution and collapse of a quasi-one-dimensional
vortex Airy beam in a Kerr medium are analyzed as a limiting case.
Our study shows the possibility of controlling and
manipulating the collapse, especially the position of collapse in the vortex Airy beam, by choosing
the initial powers, vortex orders or modulation parameters.
Analogous to optics, the results can be extended
to manage the collapse of other nonlinear waves such as fluid and matter waves.

\textit{The moments approach analysis}.
The propagation of a light beam in a Kerr medium is described in the paraxial
approximation by the following nonlinear Schrodinger (NLS) equation:
\begin{equation}
\nabla _ \bot ^2 E - 2ik\frac{{\partial E}}{{\partial z}} + \frac{{2n_2 k^2 }}{{n_0 }}\left| E \right|^2 E = 0,
\end{equation}
\noindent where $k$ is the linear wave number, $x$ and $y$ are the transverse coordinates,
$z$ is the longitudinal coordinate, $n_0$ is the linear refraction index of the medium,
$n_2$ is the third order nonlinear coefficient. Due to the complexity of the
evolution of a vortex Airy beams in a Kerr medium, we apply the approach of moments [21]
to study the nonlinear dynamics by analyzing the evolution of
several integral quantities derived from the NLS.
These quantities are defined as
\noindent
\begin{subequations}\label{eq2}
\begin{eqnarray}
I_1 (z) & = & \iint_s {\left| E \right|^2 dxdy},\\
I_2 (z) & = & \iint_s {(x^2  + y^2 )\left| E \right|^2 dxdy}, \\
I_3 (z) & = & \frac{i}
{k}\iint_s {\left[ {x\left( {E\frac{{\partial E^* }}
{{\partial x}} - E^* \frac{{\partial E}}
{{\partial x}}} \right) + y\left( {E\frac{{\partial E^* }}
{{\partial y}} - E^* \frac{{\partial E}}
{{\partial y}}} \right)} \right]}dxdy,\\
I_4 (z) & = & \frac{1}
{{2k^2 }}\iint_s {\left( {\left| {\frac{{\partial E}}
{{\partial x}}} \right|^2  + \left| {\frac{{\partial E}}
{{\partial y}}} \right|^2  - \frac{{k^2 n_2 }}
{{n_0 }}\left| E \right|^4 } \right)dxdy}.
\end{eqnarray}
\end{subequations}

These quantities are associated with the beam power $I_1$, beam width $I_2$,
momentum $I_3$, and Hamiltonian $I_4$; and satisfy a closed set of
coupled ordinary differential equations [21]: $dI_1 (z)/dz = 0$, $dI_2 (z)/dz = I_3 (z)$,
$dI_3 (z)/dz = 4I_4 (z)$, $dI_4 (z)/dz = 0$,
and the important invariant under evolution,
$Q = 2I_4 I_2  - I_3^2 /4$,
Therefore, the following  Ermakov-Pinney equation describing the
dynamics of the scaled beam width can be obtained:

\noindent
\begin{equation}\label{eq3}
\frac{{d^2 I_2^{1/2} (z)}}
{{dz^2 }} = \frac{Q}
{{I_2^{3/2} (z)}}.
\end{equation}
For a vortex Airy beam, an initial field distribution can be described by [19]:
\noindent
\begin{equation}\label{eq3}
E(x,y;z = 0) = A_0 Ai(x/x_0 )\exp (a_x x/x_0 )Ai(y/x_0 )\exp (a_y y/x_0 )(x + iy)^m,
\end{equation}
\begin{figure}
\includegraphics[height=4.3cm,width=5.5cm]{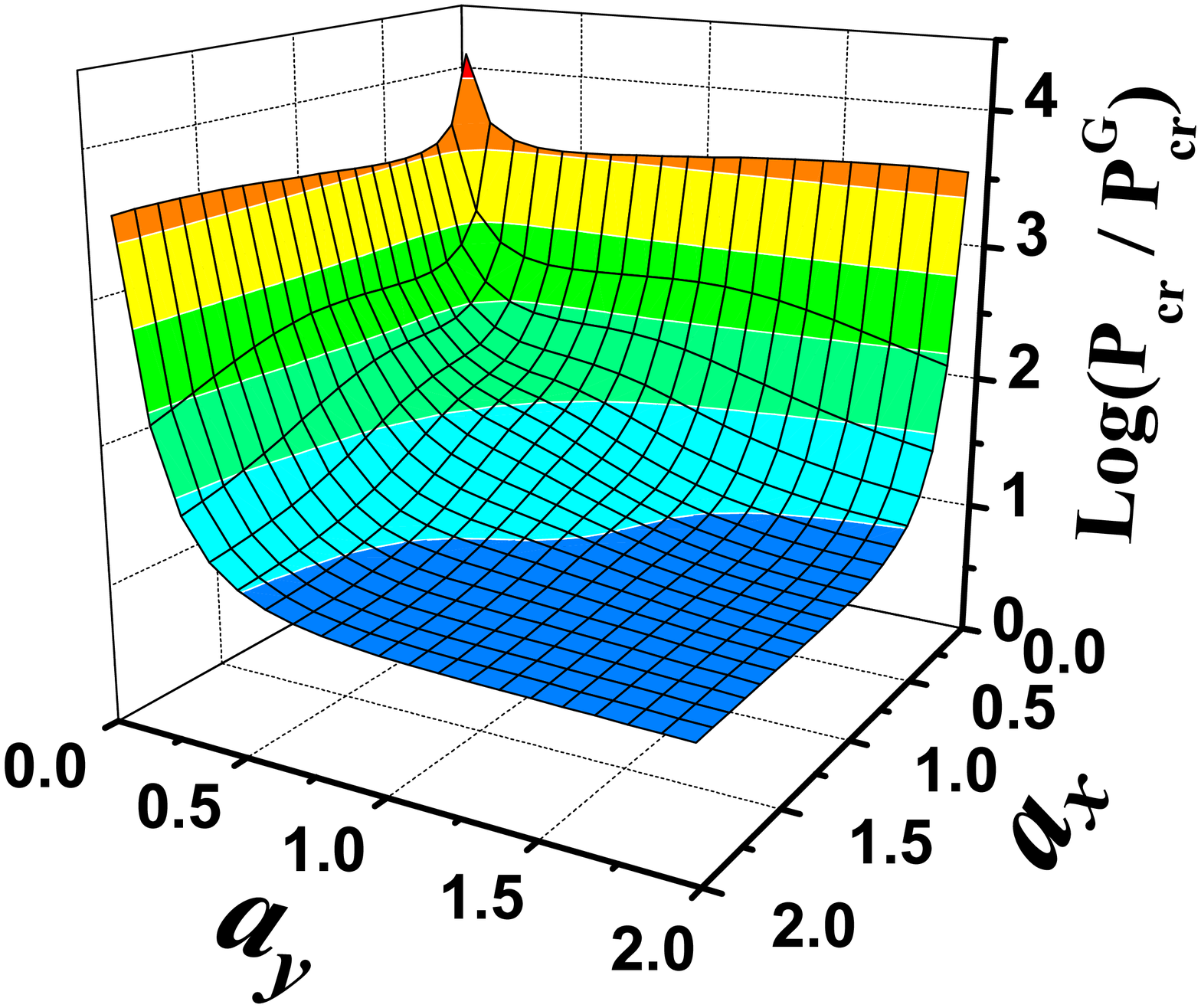}\\(a)\\[2mm]
\includegraphics[height=4.3cm,width=5.5cm]{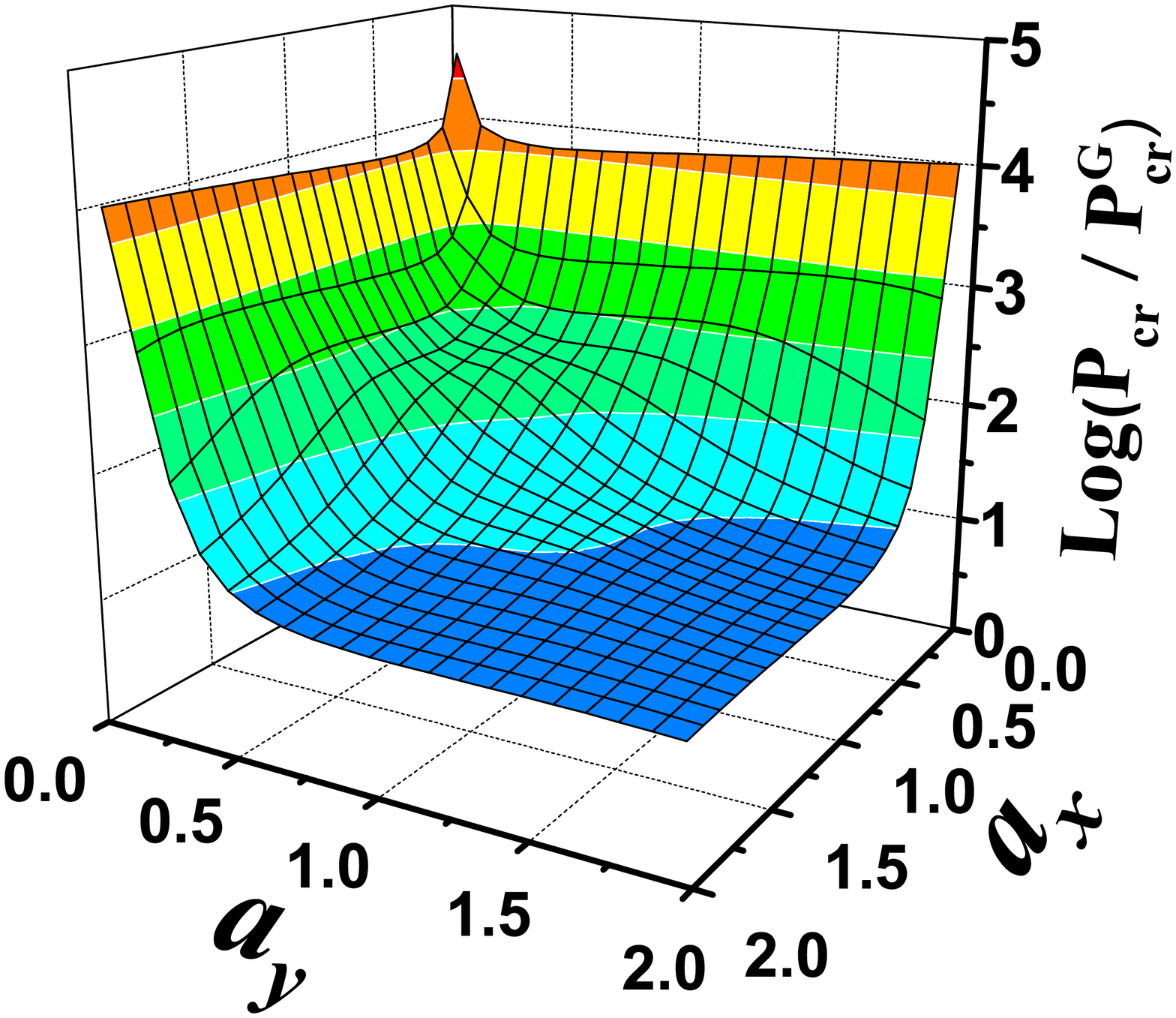}\\(b)\\
\caption{(Color online) The critical powers of vortex
Airy beams for different $a_x$ and $a_y$. (a) $m = 1$; (b) $m = 2$.}
\end{figure}
where $A_0$ is the amplitude of the complex amplitude $E (x, y, z = 0)$, $x_0$
is an arbitrary transverse scale. Exponential factors $a_x$ and $a_y$
are positive in order to
ensure containment of the infinite Airy tail in the $-x$ and $-y$ directions, respectively.
The azimuthal index $m$ represents the order of vortex.
The general solution to Eq. (3) with the vortex Airy beam as an initial
field distribution can be given as
\noindent
\begin{equation}
I_2 (z) = I_2 (z = 0) + \frac{Q}{{I_2 (z = 0)}}z^2,
\end{equation}
\noindent Eq. (5) describes the variation of the scaled beam width of
the vortex Airy beam in a Kerr medium. When $Q = 0$,
the rms beam width remains constant as recognized from Eq. (5)
and the critical value, $I_1^{cr}$, is derived from $2I_4 I_2  - I_3^2 /4 = 0$
and Eq. (2). The corresponding power can be found through
$P_{cr}  = n_0 c\varepsilon _0 I_1^{cr} /2$ [21].
The expression for $P_{cr}$ is not presented here due to its lengthy expression.

Figure 1 shows the denary logarithmic vertical scale of the ratio of
the critical power of a vortex Airy beam (with $m = 1, 2$)
to the critical power $P_{cr}^G  = \pi c\varepsilon _0 n_0^2 /(n_2 k^2 )$
of a Gaussian beam ,
where $c$ is the speed of light in vacuum and
$\varepsilon _0$ is the permittivity of free space.
In Fig. 1 one sees that the value of the critical power
of the vortex Airy beam depends mainly on the beam profile of
the transverse distribution (besides the nonlinear parameters of the medium).
The critical power of the vortex Airy beam increases as the the order of vortex modes
increases, but decreases as the modulation parameters $a_x$ and $a_y$ increase.
If the initial power exceeds the critical power $P_{cr}$,
the rms beam width goes to zero at a finite propagation distance,
as predicted by the approach of moments. Obviously, the critical power
is the upper bound for the collapse power [22, 23].

\textit{Evolution and collapse of vortex Airy beams in Kerr media}.
Let us examine the evolution of a vortex Airy beam in a Kerr medium.
In the numerical calculation,
we take $\lambda= 0.53\mu m$, $x_0 = 100\mu m$ and $z_0  = kx_0^2 /2 = 6cm$.
The intensity distributions of the 1st and 2nd vortex Airy beams
with $P_{in}  = 5P_{cr}^G$ and $P_{in}  = 15P_{cr}^G$ in
the focusing nonlinear medium at different propagation distances are shown in Fig. 2,
where hereafter the intensity distribution is normalized with respect
to its initial peak intensity. One sees that the beams with different
initial powers in the Kerr medium propagate along the same accelerating
curve trajectory as an Airy beam in free space, as expected.
The collapse will not occur during the propagation for these cases but
the evolution of intensity distributions of the beam,
however, depends sensitively on the order of vortex and the initial power.
\begin{figure}
\includegraphics[height=6cm,width=7.5cm]{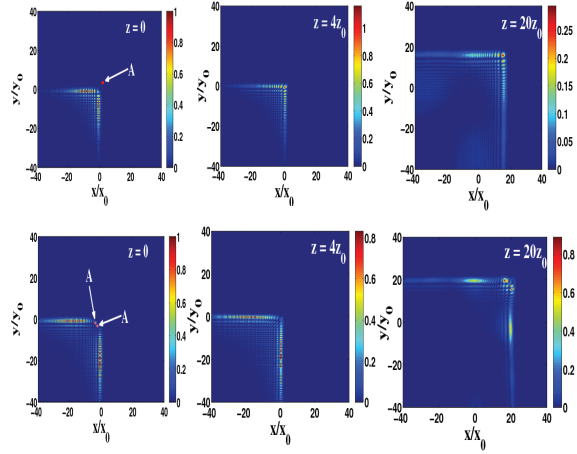}
\caption{(Color online)   (Color online) The intensity distribution of
 vortex Airy beam $(a_x = a_y = 0.1)$ at different
propagation distances with initial powers  in focusing Kerr medium.
Upper row:the $m = 1$, $P_{in} = 5P_{cr}^G=0.0023P_{cr}$;  and lower: $m = 2$,
$P_{in} = 15P_{cr}^G=0.0024P_{cr}$. }
\end{figure}

\begin{figure}
\includegraphics[height=7cm,width=9cm]{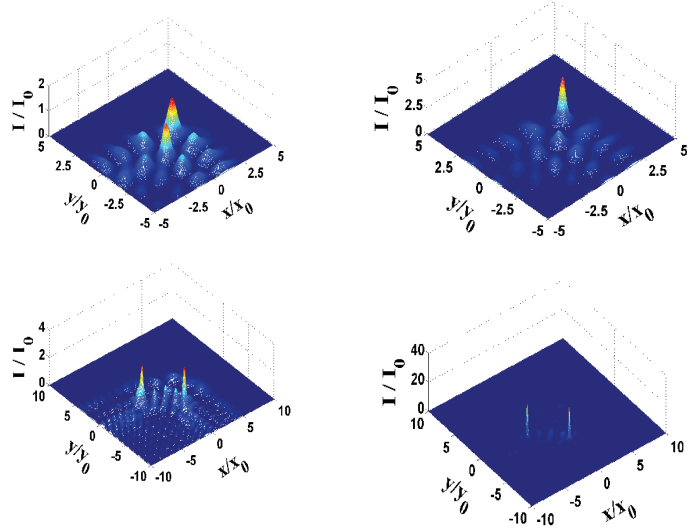}
\caption{(Color online) The intensity distribution of the vortex Airy beam $(a_x = a_y = 0.1)$
in focusing medium.
(a) $m=1$,$P_{in} = 15P_{cr}^G=0.0069P_{cr}$, $z = 5z_0$;
(b)$m=1$, $P_{in} = 15P_{cr}^G=0.0069P_{cr}$, $z = 6z_0$;
(c) $m=2$,$P_{in} = 50P_{cr}^G=0.0079P_{cr}$, $z = 7z_0$;
(d) $m=2$,$P_{in} = 50P_{cr}^G=0.0079P_{cr}$, $z = 7.3z_0$;}
\end{figure}
\begin{figure}
\includegraphics[height=7cm,width=9cm]{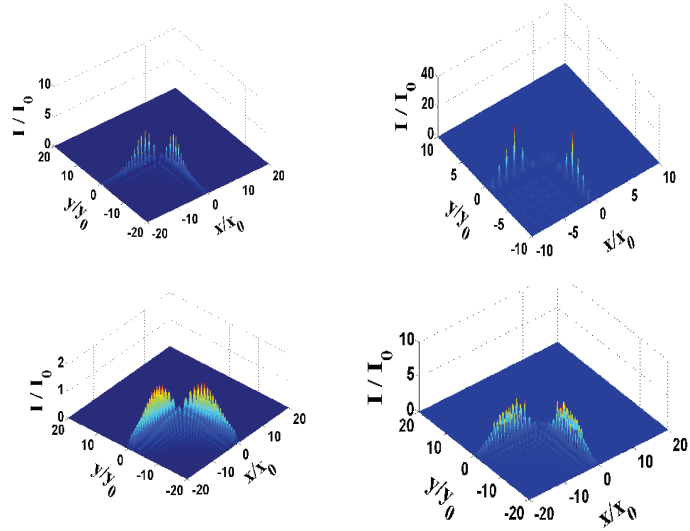}
\caption{(Color online) The intensity distribution of the vortex Airy beam $(a_x = a_y = 0.1)$
in focusing medium.
(a) $m=1$,$P_{in} = 50P_{cr}^G=0.023P_{cr}$, $z = 0.7z_0$;
(b) $m=1$,$P_{in} = 50P_{cr}^G=0.023P_{cr}$,$z = 0.8z_0$.
(c) $m=1$,$P_{in} = 2500P_{cr}^G=1.15P_{cr}$, $z = 0.02z_0$;
(d) $m=1$,$P_{in} = 2500P_{cr}^G=1.15P_{cr}$,$z = 0.03z_0$;}
\end{figure}
\begin{figure}
\includegraphics[height=4.3cm,width=5.5cm]{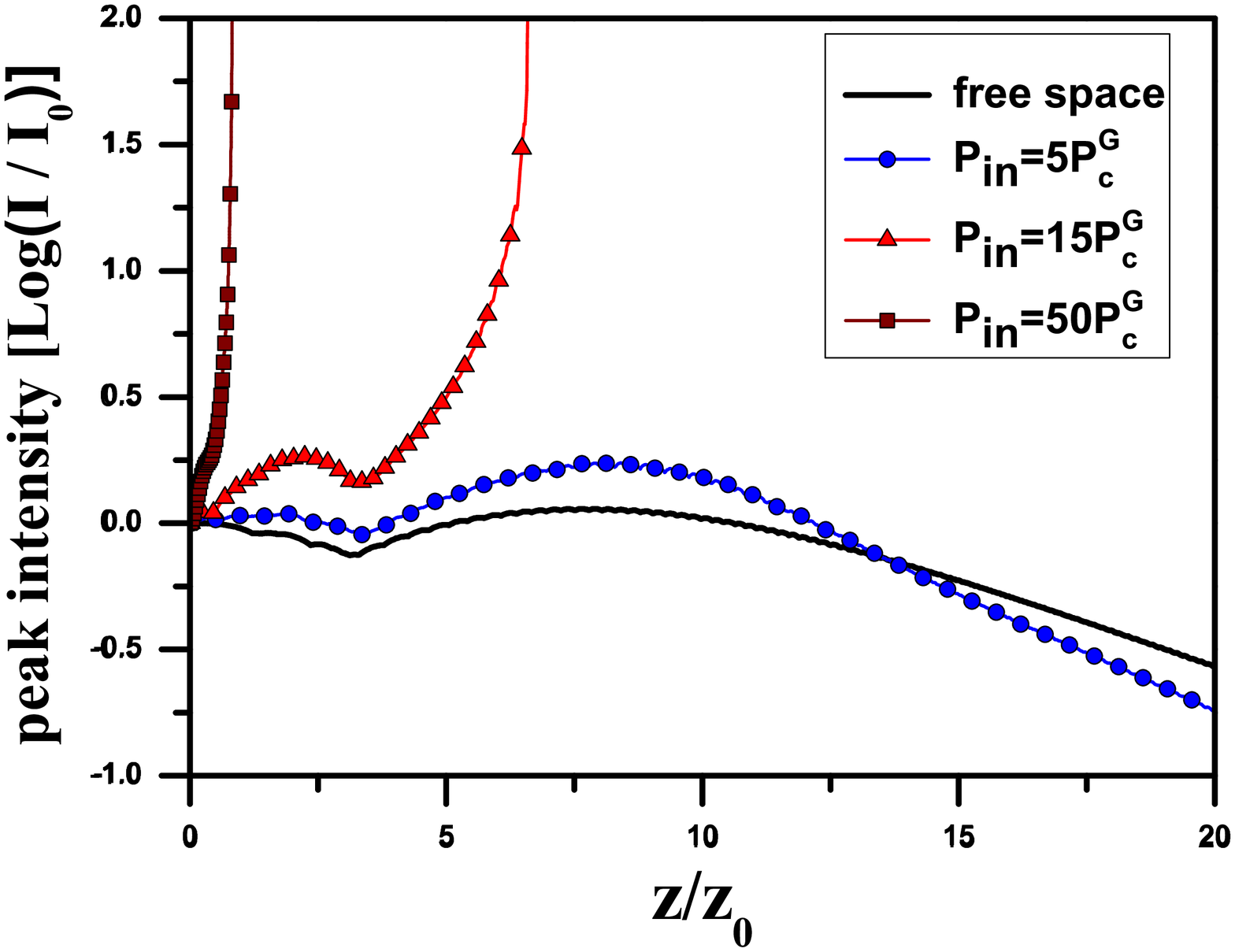} \\(a)\\[2mm]
\includegraphics[height=4.3cm,width=5.5cm]{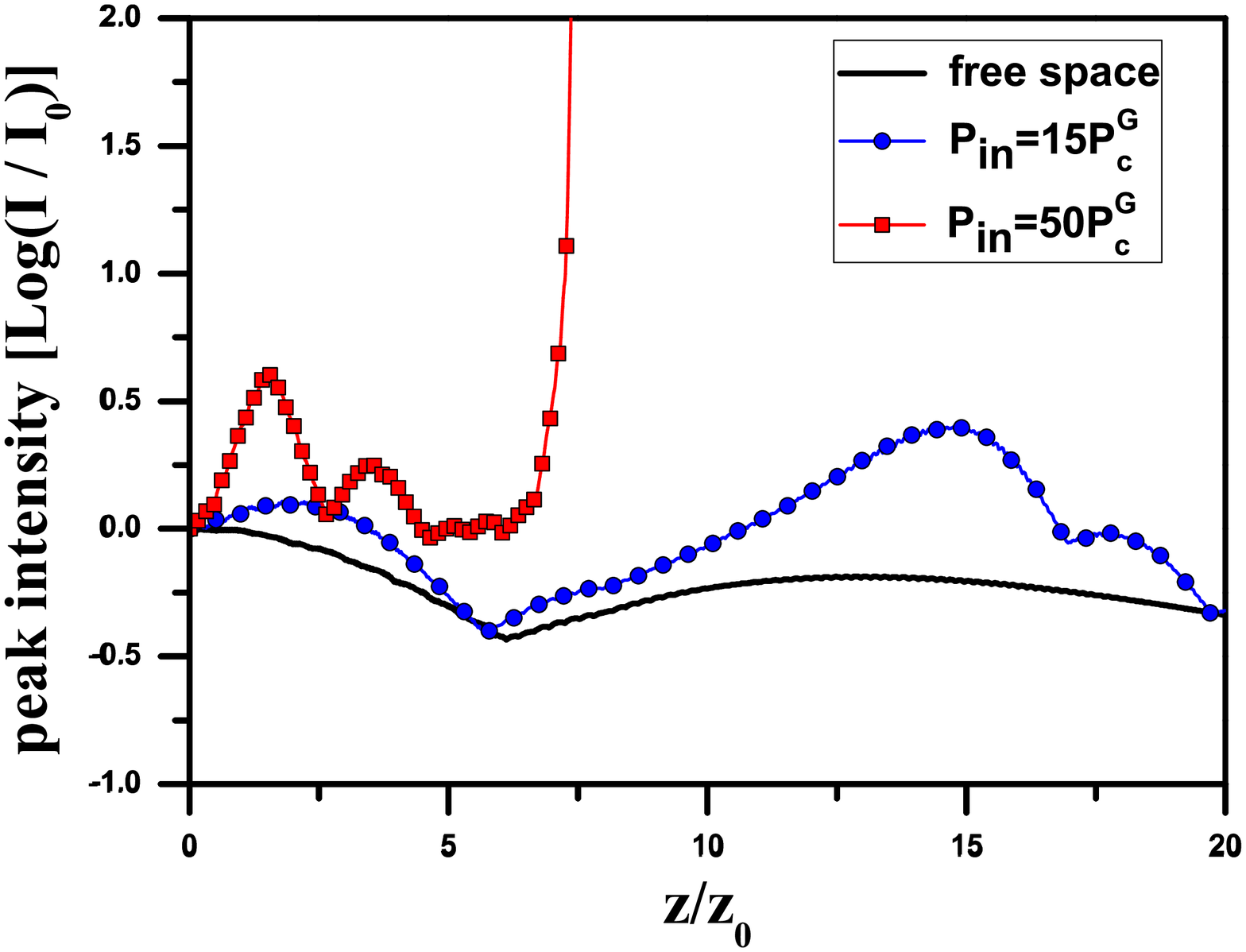} \\(b)
\caption{(Color online) The peak intensity of vortex Airy beam
$a_x = a_y = 0.1$ as a function of the propagation distance with
different initial powers (a) $m = 1$; (b) $m = 2$.}
\end{figure}
As the initial power increases, more energy accumulates at various
locations of the beam, leading to partial collapse of the beam.
If the initial power is $P_{in}  = 15P_{cr}^G$, the 1st order vortex Airy beam
collapses at the outermost main lobe of the beam, as illustrated in
Fig. 3 (upper).  No collapse of vortex Airy beam with the order of $m = 2$
is expected, if $P_{in}  = 15P_{cr}^G$ (see Fig. 2 (lower)).
The 2nd order vortex Airy beam, however, is expected to collapse at
$P_{in}  = 50P_{cr}^G$, as shown in Fig. 3(c) and 3(d). In this case, the partial collapse
will occur separately in two lobes in the beam (but, not in the outermost lobes).
Numerical simulations indicate that the actual collapse power for
the 1st order vortex Airy beam with $a_x = a_y = 0.1$ is $P_{in} = 0.0063P_{cr} = 14P^G_{cr}$
and that for the 2nd order vortex Airy beam with $a_x = a_y = 0.1$ is
$P_{in} = 0.0058P_{cr} = 37P^G_{cr}$.
It is interesting to see that
the 1st order vortex Airy beam collapses at the outermost lobes,
if $P_{in} = 50P_{cr}^G$ (see Fig. 4(a) and (b)).
A stronger initial power increases the number of lobes that
collapse at the outermost region, and decreases the propagation
distances of beam (e.g. Fig. 4(c) and 4(d) with $P_{in} = 1.15P_{cr} = 2500P^G_{cr}$).
In Fig. 5, we show the normalized peak intensities as a function of
the propagation distance with different initial powers. Although
the approach of moments predicts that the rms beam width will be
broadened when $P_{in}<P_{cr}$,
the results from the variation of peak intensities suggest
a deformation and redistribution of the vortex Airy beams during the propagation.
The intensity at some lobes will dominate and eventually collapse
while the rms beam width increases or remains constant, such as for the cases
$P_{in}  = 15P_{cr}^G$ and $50P_{cr}^G$ for the 1st order vortex Airy beam
and $P_{in}  = 50P_{cr}^G$ for the 2nd order vortex Airy beam.
The results indicate that a higher initial power leads a shorter propagation distance before collapse.
When the initial power begin to exceed the collapse power,
the flow of transverse energy of Airy beam and vortex
leads to a collapse occurring at the outermost main lobe of the beam.
It is interesting to see that the energy accumulates to the position where the field
is almost zero and finally a collapse occurs.
On the other hand, no collapse occurs at a position where initial field mainly exists.
In all the studies reported in the literature (for either vortex or non-vortex beams),
the field never collapses at a location where the initial field is almost zero [3,22,23].
In the present study, however, a unusual collapse is observed for the present vortex Airy beam.
The vortex Airy beam may collapse at a location where the initial
field is nearly zero, e.g., the collapse point in Fig. 3(b)
(cf. the zero initial field at this point, marked as point A on the first subfigure of Fig.2).
During the propagation, the energy accumulates at this collapse position.
The higher the initial power value, the closer the collapse position to the beam's
center and the shorter propagation distance before collapse.
These results, therefore, provide useful information on how to spatially manipulate a collapse in experiment.
If the initial power increases further to a certain threshold value,
the beam collapses at many side lobes simultaneously, but not in the outermost lobes,
see e.g. the collapse position in Fig. 3(d)
(cf. the initial field at these points around point A in Fig. 2 with $z = 0$).
In this case, we see that the position of collapse, the number of collapse lobes,
and the propagation distance (before collapse) also depend on the initial power.
As the initial power reaches a considerably high value,
the collapse will appear at many outermost lobes with a very short
propagation distance with a higher initial power and a larger number of collapsed lobes.
Even if the approach of moments
predicts that the rms beam width will go to be zero at some propagation
distance when the initial power exceeds the critical power $P_{cr}$,
the vortex Airy beam collapses separately at the outermost lobes as shown in Fig. 4(c) and 4(d).

\textit{Quasi-one-dimensional vortex Airy beam}.
\noindent We now investigate the effect of modulation parameters on the nonlinear
evolution of the vortex Airy beam. When $a_x=1.5$ and $a_y=0.05$,
the beam becomes a quasi-one-dimensional vortex Airy beam, as shown in Fig. 6.
Under a low initial power of $P_{in}  = 1.5P_{cr}^G$,
the beam propagates along the accelerating trajectory with a feature similar to
the propagation behavior of an Airy beam in free space,
the field distribution is evidently different with that of the beam in free space,
and the beam does not collapse as shown in Fig. 6(a). By increasing the initial power from
$P_{in}  = 1.5P_{cr}^G$ to $5.5P_{cr}^G$, the off-center part of the beam
partially collapses at certain propagation distance although the propagation
trajectory is the same as that of $P_{in}  = 1.5P_{cr}^G$,
(compared with the zero initial field of this collapse point to the point A on first subfigure in Fig. 6(b)).
On the other hand, no collapse occurs at the position where the field originally exists as shown in Fig. 6(b).
The underlying physics are similar to the collapse behavior of a vortex Airy beam.

\begin{figure}
\includegraphics*[height=7cm,width=9cm]{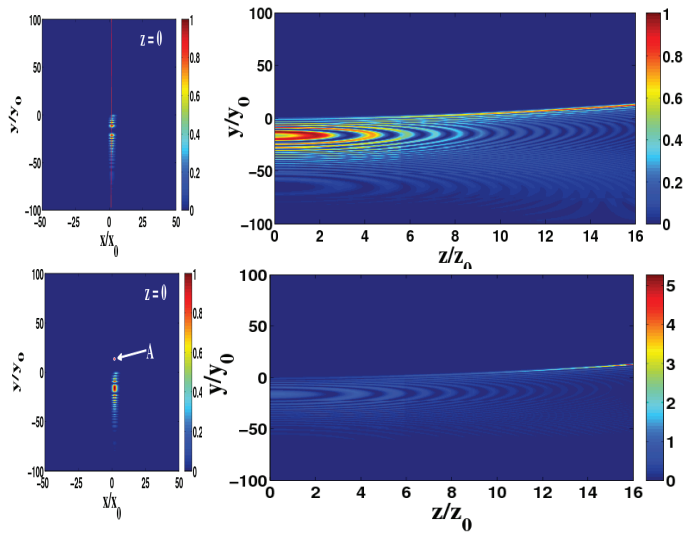}
\caption{(Color online) The intensity distribution of a vortex Airy beam
($a_x = 1.5, a_y =0.05$) at different propagation distances with
initial powers (a)$P_{in} = 1.5P_{cr}^G = 0.00043P_{cr}$; (b) $P_{in} = 5.5P_{cr}^G= 0.00157P_{cr}$.
The thin red line in the first
plot indicates the position of the longitudinal cross-section.}
\end{figure}
%
In conclusion, we have systematically studied the
collapse dynamics of vortex Airy optical fields in a Kerr medium.
The collapse of a vortex Airy beam requires a higher
energy than that of a non-vortex beam. Although vortex
Airy beams with different initial powers propagate along a similar
acceleration curve trajectory in the Kerr medium as a non-vortex Airy
beam propagates in free space,
the evolution and appearance of the collapse in a vortex Airy beam exhibits
unusual features.
The collapse can occur in the position where the initial field is almost zero.
Our study shows that the collapse can occur at different
locations other than the central point,
and the locations can be controlled through choosing appropriate initial powers, the order of vortex,
and the modulation parameters.
This work is supported in part by the State Key Program for Basic Research of China
under Grant No.2006CB921805, the Science and Technology Department of Zhejiang Province (2010R50007),
and the Key Project of the Education
Commission of Zhejiang Province of China No.Z201120128.


\end{document}